\documentclass[letter,desactivate]{aa} 

\usepackage{natbib}
\bibpunct{(}{)}{;}{a}{}{,}

\usepackage{graphicx}
\usepackage{revsymb}	
\usepackage{amsmath}
\usepackage{txfonts}
\usepackage{amssymb}
\usepackage{geometry} 
\usepackage[parfill]{parskip} 
\usepackage{slantsc}
\usepackage{array}
\usepackage{amsfonts}
	
\usepackage{epstopdf}	
\usepackage{epsfig}	

\usepackage{url}

\usepackage[colorlinks=true,linkcolor=black, urlcolor=black, citecolor=black]{hyperref}

\usepackage{color}

\usepackage{xfrac}

\usepackage{sidecap}
\usepackage[font=small, labelfont=bf]{caption}
\usepackage{subcaption}

\usepackage{threeparttable}

\begin{document}
   \title{Imprint of the magnetic activity cycle on solar asteroseismic characterisation based on 26 years of GOLF and BiSON data}
\author{J. B\'{e}trisey\inst{1,2} \and M. Farnir\inst{3} \and S.~N. Breton\inst{4} \and R.~A. Garc{\'\i}a\inst{5} \and A. -M. Broomhall\inst{6,7} \and A.~M. Amarsi\inst{2} \and O. Kochukhov\inst{2}}
\institute{Observatoire de Genève, Université de Genève, Chemin Pegasi 51, 1290 Versoix, Suisse\\email: 	\texttt{Jerome.Betrisey@unige.ch}
\and Department of Physics and Astronomy, Uppsala University, Box 516, SE-751 20 Uppsala, Sweden
\and STAR Institute, University of Liège, 19C Allée du 6 Août, 4000 Liège, Belgium
\and  INAF – Osservatorio Astrofisico di Catania, Via S. Sofia 78, 95123 Catania, Italy
\and Université Paris-Saclay, Université Paris Cité, CEA, CNRS, AIM, 91191 Gif-sur-Yvette, France
\and Centre for Fusion, Space and Astrophysics, Department of Physics, University of Warwick, Coventry CV4 7AL, UK
\and Centre for Exoplanets and Habitability, University of Warwick, Coventry CV4 7AL, UK}
\date{\today}

\abstract
{Asteroseismic modelling will play a key role in future space-based missions, such as PLATO, CubeSpec, and Roman. Despite remarkable achievements, asteroseismology has revealed significant discrepancies between observations and theoretical predictions of the physics used in stellar models, which have the potential to bias stellar characterisation at the precision level demanded by PLATO. The current modelling strategies largely overlook magnetic activity, assuming that its effects are masked within the parameterisation of the so-called `surface effects'. Given the presence of activity cycles in multiple solar-like oscillators and activity variations in a significant fraction of \textit{Kepler} observations of main-sequence stars, it is therefore relevant to investigate systematic errors in asteroseismic characterisations caused by our incomplete understanding of magnetic activity.}
{Based on 26 years of GOLF and BiSON observations, we measured the impact of magnetic activity on the asteroseismic characterisation of the Sun-as-a-star, a reference target for assessing the PLATO mission requirements.}
{The GOLF and BiSON observations, which fully cover solar cycles 23 and 24, were divided into yearly overlapping snapshots, each delayed by a quarter of a year. For each snapshot, an advanced asteroseismic characterisation, similar to that to be adopted by the PLATO pipeline, was performed with standard prescriptions for the parameterisation of the surface effects. This allowed for the apparent temporal evolution of fundamental solar parameters such as mass, radius, age to be monitored. The correlation of these parameters with the 10.7 cm radio emission flux, a proxy of the solar activity cycle, was then measured.}
{The effects of magnetic activity are partially absorbed into the parameterisation of the surface effects when suitable
prescriptions are used, and do not significantly affect the measured solar mass or radius. However, contrary to literature expectations, we find a significant imprint on the age determination, with variations of up to 6.5\% between solar minima and maxima. This imprint persists across both BiSON and GOLF datasets.}
{Considering that the Sun exhibits low levels of activity, our study highlights the looming challenge posed by magnetic activity for future photometry missions, and prompts a potential reevaluation of the asteroseismic characterisation of \textit{Kepler}’s most active targets.}


\maketitle

\defcitealias{Betrisey2023_AMS_surf}{JB23}

\section{Introduction}
Convective motions in the upper layers of solar-type stars excite a broad spectrum of stellar oscillations. Through the study of these oscillations, asteroseismology allows us to probe the internal structure of stars and determine their key parameters, such as mass, radius, and age, with a precision and accuracy unmatched by other standard techniques for non-binary stars. Precise and accurate stellar models are crucial for understanding planetary system evolution and unravelling the history of our own galaxy through Galactic Archaeology \citep[see e.g.][]{Chaplin&Miglio2013,Garcia&Ballot2019}. Building on the success of previous missions such as CoRoT \citep{Baglin2009}, \textit{Kepler} \citep{Borucki2010}, K2 \citep{Howell2014}, and TESS \citep{Ricker2015}, asteroseismic modelling will play a key role in the future PLATO \citep{Rauer2024}, CubeSpec \citep{Bowman2022}, and Roman \citep{Huber2023} space-based missions. 

Asteroseismology has also highlighted significant discrepancies between observations and theoretical predictions of the physics used in stellar models that can bias stellar characterisation, especially with the precision required by the PLATO mission (15\% in mass, 1-2\% in radius, and 10\% for a Sun-like star). In particular, the treatment of near-surface layers \citep[e.g.][]{Ball&Gizon2017,Nsamba2018,Jorgensen2020,Jorgensen2021,
Cunha2021,Betrisey2023_AMS_surf} and the choice of physical ingredients in stellar models \citep[e.g.][]{Buldgen2019f,Farnir2020,Betrisey2022} pose substantial challenges. Modelling inaccuracies in the near-surface layers, known in the literature as surface effects, arise notably from the inaccurate treatment of convection in 1D stellar evolutionary models and from neglecting non-adiabatic effects in the oscillation code. As a result, surface effects induce frequency shifts that are frequency-dependent, with respect to eigenfrequencies computed with 1D adiabatic oscillation codes \citep[see e.g.][]{Kjeldsen2008}.

However, magnetic stellar activity can also alter the observed frequencies, complicating the situation further. The current literature on magnetic activity is incomplete, and distinguishing between magnetic activity and surface effects remains unclear. From a theoretical standpoint, frequency shifts due to magnetic activity might originate from structural variations of the sub-surface layers \citep[e.g.][]{Woodard&Noyes1985,Fossat1987,Libbrecht&Woodard1990,Kuhn1998,Dziembowski&Goode2005,Basu2012} and magnetic fields \citep[e.g.][]{Howe2002, Baldner2009}. In the state of the art, the current modelling strategies largely overlook magnetic activity, assuming that its effects are masked within the parameterisation of the surface effects \citep[see e.g.][and references therein]{PerezHernandez2019}. Recent studies, however, challenged this perspective. \citet{PerezHernandez2019} showed that magnetic activity can have a small but non-negligible impact on the estimation of stellar mass, radius, age, and helium abundance in two main-sequence stars. \citet{Thomas2021} further showed, using artificial data, that magnetic activity might introduce substantial biases in stellar parameter estimation, at the precision level required by the PLATO mission. These two studies reported biases up to 10\% and 5\% for the estimation of the stellar age, respectively, which are comparable to the 10\% of precision in age mandated by PLATO. Considering that activity cycles have been detected in multiple solar-like oscillators \citep[e.g.][]{Garica2010,Regulo2016,Salabert2016,Kiefer2017,Salabert2018,Santos2018,Santos2019_sig} and activity variations were found in a significant fraction of \textit{Kepler} observations of main-sequence stars \citep[e.g.][]{Santos2019_rot,Santos2021,Santos2023}, it becomes essential to explore and quantify systematically the influence of magnetic activity on stellar characterisation in preparation for future space-based missions.

To better understand the impact of stellar activity, we decided to examine solar data. The Sun, as the nearest and most extensively studied star, indeed serves as an ideal laboratory for measuring the effects of stellar activity. It is the only star with decades of continuous observations of acoustic oscillations, covering several activity cycles (see Appendix \ref{app:solar_observational_context}). Solar observations have revealed that low-degree acoustic frequencies change with the 11-year solar activity cycle \citep{Woodard&Noyes1985}, a finding then verified by numerous studies for low and intermediate modes \citep[see e.g.][]{Broomhall&Nakariakov2015}. Additionally, quasi-biennial oscillations have been detected in solar data \citep[see e.g.][]{Mehta2022}, though their physical origin remains unclear \citep[see e.g.][]{Bazilevskaya2014}. Moreover, the imprint of the solar activity cycle has been observed in global seismic observables such as the large separation \citep{Broomhall2011} and the frequency of maximum power \citep{Howe2020}. Magnetic activity therefore directly impacts stellar characterisation when scaling relations are employed \citep[see e.g.][for a review about scaling relations]{Hekker2020}. Please refer to Sect. 3.2.4 of \citet{Betrisey2024_phd} for a more compete literature review of the impact of magnetic activity on solar acoustic frequencies.

In this letter, we investigate the influence of magnetic activity on the asteroseismic characterisation of the Sun-as-a-star, utilising an advanced `à la PLATO' modelling approach. Our analysis is based on 26 years of data from GOLF and BiSON Doppler velocity observations. In Sect.~\ref{sec:modelling_procedure}, we detail the datasets and outline the modelling strategy employed for the characterisation. In Sect.~\ref{sec:results}, we assess the correlation between various stellar parameters derived from the characterisation and the 10.7 cm radio emission flux, which serves as a proxy of the solar activity cycle. Finally, in Sect.~\ref{sec:conclusions}, we present our conclusions.

\begin{table}[t]
\caption{Pearson correlation coefficient between the solar asteroseismic age and the solar activity cycle proxy, the 10.7 cm radio emission flux.}
\centering
\begin{tabular}{lccc}
\hline \hline
 & cycle 23 & cycle 24 & two cycles \\ 
\hline 
\textit{GOLF} &  &  &  \\ 
BG2, $n\geq 12$ & $0.57\pm 0.12$ & $0.43\pm 0.12$ & $0.50\pm 0.10$ \\
BG2, $n\geq 16$ & $0.66\pm 0.08$ & $0.61\pm 0.09$ & $0.60\pm 0.05$ \\
BG2, $n\geq 18$ &  $0.14\pm 0.27$ &  $0.27\pm 0.19$ &  $0.25\pm 0.13$ \\
K1, $n\geq 12$ & $0.69\pm 0.10$ & $0.32\pm 0.14$ & $0.40\pm 0.02$ \\
K1, $n\geq 16$ & $0.76\pm 0.05$ & $0.35\pm 0.05$ & $0.57\pm 0.09$ \\
S1, $n\geq 12$ &  \multicolumn{3}{c}{MCMC did not converge} \\
S1, $n\geq 16$ &  \multicolumn{3}{c}{MCMC did not converge} \\
\hline 
\textit{BiSON} &  &  &  \\ 
BG2, $n\geq 12$ & $0.64\pm 0.13$ & $0.64\pm 0.11$ & $0.65\pm 0.12$ \\
BG2, $n\geq 16$ & $0.66\pm 0.06$ & $0.58\pm 0.07$ & $0.67\pm 0.05$ \\
\hline
\end{tabular} 
\label{tab:pearsonr_age}
\end{table}

\section{Datasets and modelling procedure}
\label{sec:modelling_procedure}
Our observational data is composed of high-quality Sun-as-a-star measurements of pressure modes across solar cycles 23 and 24. To ensure robust detection of correlations with the solar activity cycle, we based our study on two independent datasets: GOLF \citep{Gabriel1995} and BiSON \citep{Davies2014_bison,Hale2016} observations. GOLF, monitoring the Sun from space, and BiSON, observing from the ground, are both sensitive to radial velocity variations, enabling the extraction of high-quality pressure modes. The GOLF observations were divided into 94 yearly overlapping snapshots, each delayed by 91.25 days. Similarly, the BiSON observations were divided into 92 overlapping snapshots. The detailed modelling procedure for acoustic oscillation extraction is provided in Appendix~\ref{app:detailed_modeling_procedure}.

For each snapshot, we characterised the solar parameters using an advanced modelling procedure similar to that to be
adopted by the PLATO pipeline. This involves fitting acoustic frequencies and non-seismic constraints (in our case, the spectroscopic constraints: effective temperature, metallicity, and luminosity) using the MCMC-based AIMS software \citep{Rendle2019}. We refer to Appendix~\ref{app:detailed_modeling_procedure} for a detailed description of the modelling strategy. The uncertainties of the non-seismic constraints were adjusted to match the data quality of the best \textit{Kepler} targets. For our study, we used the standard MS subgrid of the \textit{Spelaion} grid \citep[][hereafter JB23]{Betrisey2023_AMS_surf}. The combination of this high-resolution grid and the interpolation scheme of AIMS allows for thorough exploration of the parameter space. We optimised four main free parameters (mass, age, and initial hydrogen and helium mass fractions $X_0$ and $Y_0$), along with one or two additional free parameters depending on the surface effect prescription considered.

According to the literature, the impact of magnetic activity should be masked within the parameterisation of the surface effects \citep[see e.g.][and references therein]{PerezHernandez2019}. However, this has primarily been studied using the \citet{Ball&Gizon2014} surface effect prescription \citep{Howe2017}. Thus, we also examined the two other main prescriptions from the literature \citep{Kjeldsen2008,Sonoi2015}. It should be noted that if frequency shifts due to magnetic activity are not monotonically increasing with frequency like in the Sun, this indirect treatment of magnetic activity might not be effective \citep{Salabert2018}. Similarly to surface effects, the impact of magnetic activity is stronger on higher-order oscillation frequencies. To investigate whether the characterisation based on mode sets composed of higher-order frequencies is more likely to be affected by magnetic activity, we tested different mode sets by gradually removing the lowest-order modes. For the Sun, instruments like GOLF and BiSON can detect lower-order frequencies more effectively than the VIRGO \citep{Frohlich1995} instrument due to differences in photometric background. \textit{Kepler} exhibits a similar behaviour to VIRGO, and this is expected to be the case for PLATO as well. Consequently, PLATO observations may potentially be more sensitive to the effects of magnetic activity. The different configurations investigated in our study are summarised in Table~\ref{tab:pearsonr_age}. We used the same abbreviations as in \citetalias{Betrisey2023_AMS_surf} for the surface effect prescriptions: BG2 for the two-term \citet{Ball&Gizon2014} prescription, S1 for the one-term \citet{Sonoi2015} prescription, and K1 for the one-term \citet{Kjeldsen2008} prescription. As discussed in \citetalias{Betrisey2023_AMS_surf} for example, the two-term variants of the S1 and K1 prescriptions are unsuitable for asteroseismic targets due to a non-linear free coefficient that destabilises the minimisation procedure, preventing successful convergence for most cases outside of solar conditions.

\section{Imprint of the magnetic activity cycle}
\label{sec:results}
As a well-established proxy for solar activity \citep[see e.g.][and references therein]{Tapping2013}, we used the 10.7 cm radio emission flux\footnote{see \url{https://www.spaceweather.gc.ca/}}. The correlation between solar parameters and the magnetic activity cycle is then evaluated by computing the Pearson correlation coefficient $\mathcal{R}$ \citep{Pearson1895} between the solar parameters and the 10.7 cm radio emission flux. We refer to Appendix~\ref{app:supplementary_data} for a detailed description of the evaluation procedure of the Pearson coefficient.

\begin{figure*}
\resizebox{0.99\linewidth}{!}{
\begin{subfigure}[b]{.33\textwidth}
  \includegraphics[width=.99\linewidth]{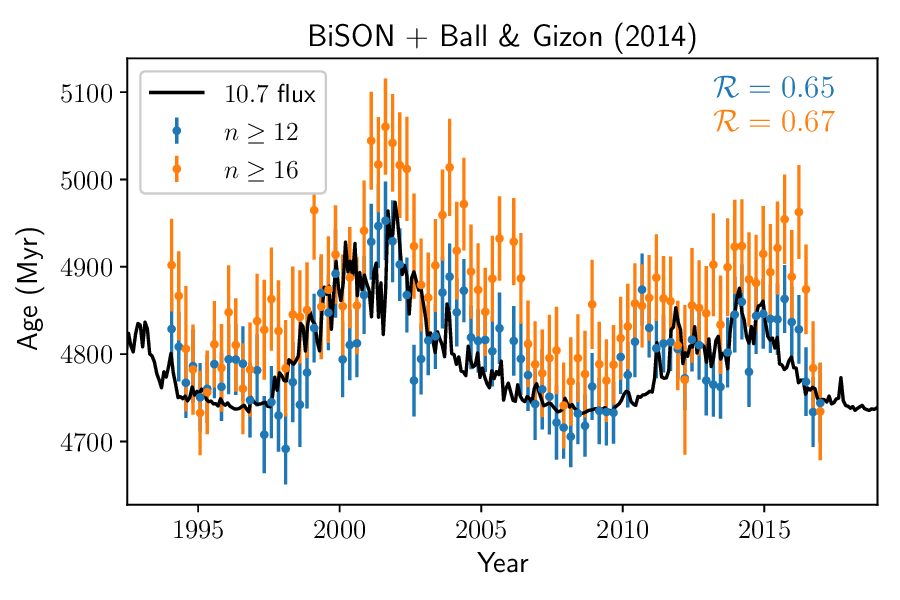}  
\end{subfigure}
\begin{subfigure}[b]{.33\textwidth}
  \includegraphics[width=.99\linewidth]{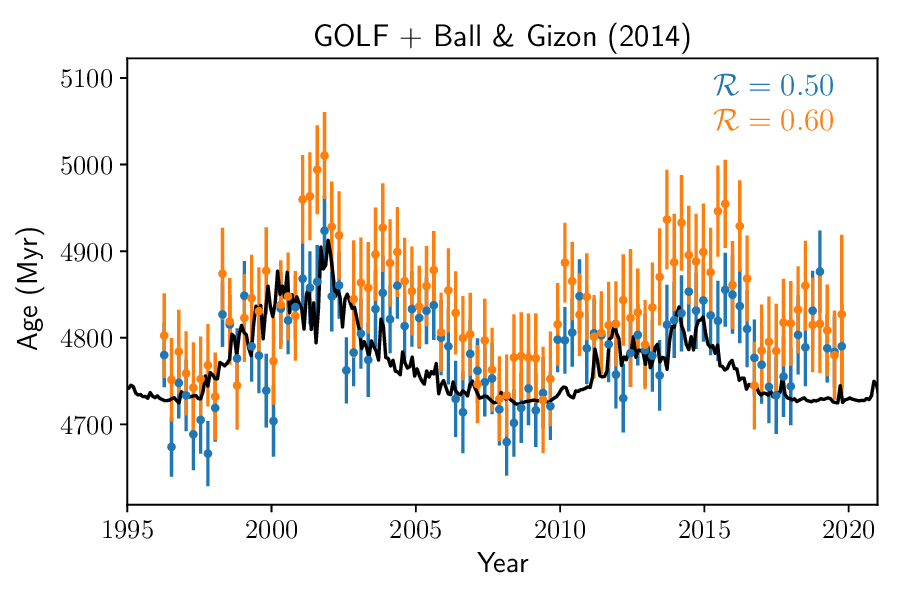} 
\end{subfigure}
\begin{subfigure}[b]{.33\textwidth}
  \includegraphics[width=.99\linewidth]{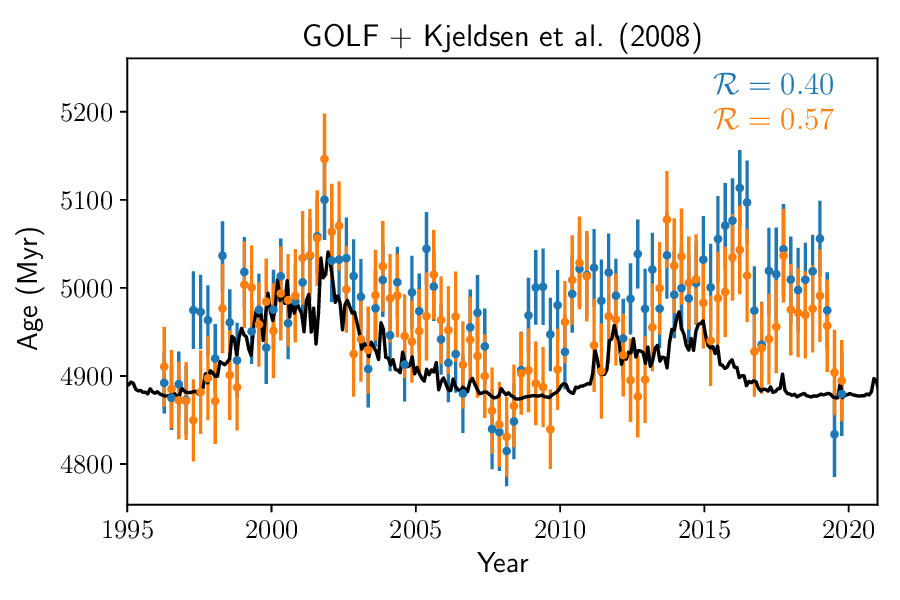} 
\end{subfigure}}
\resizebox{0.99\linewidth}{!}{
\begin{subfigure}[b]{.33\textwidth}
  \includegraphics[width=.99\linewidth]{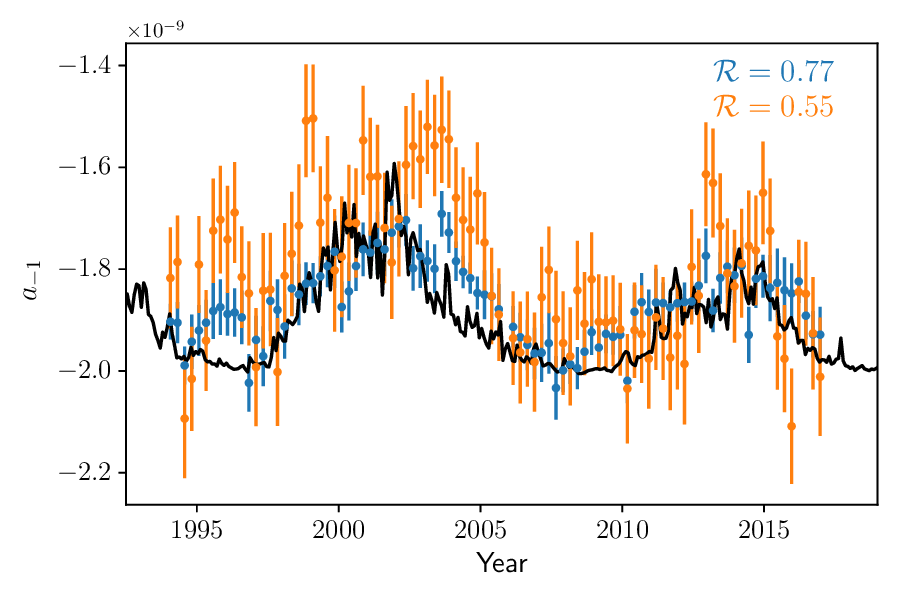}  
\end{subfigure}
\begin{subfigure}[b]{.33\textwidth}
  \includegraphics[width=.99\linewidth]{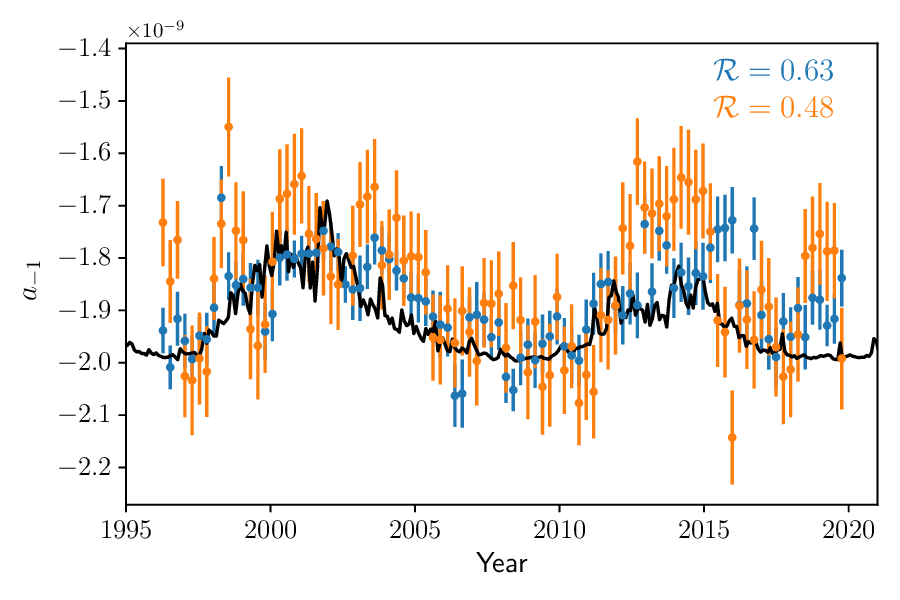} 
\end{subfigure}
\begin{subfigure}[b]{.33\textwidth}
  \includegraphics[width=.99\linewidth]{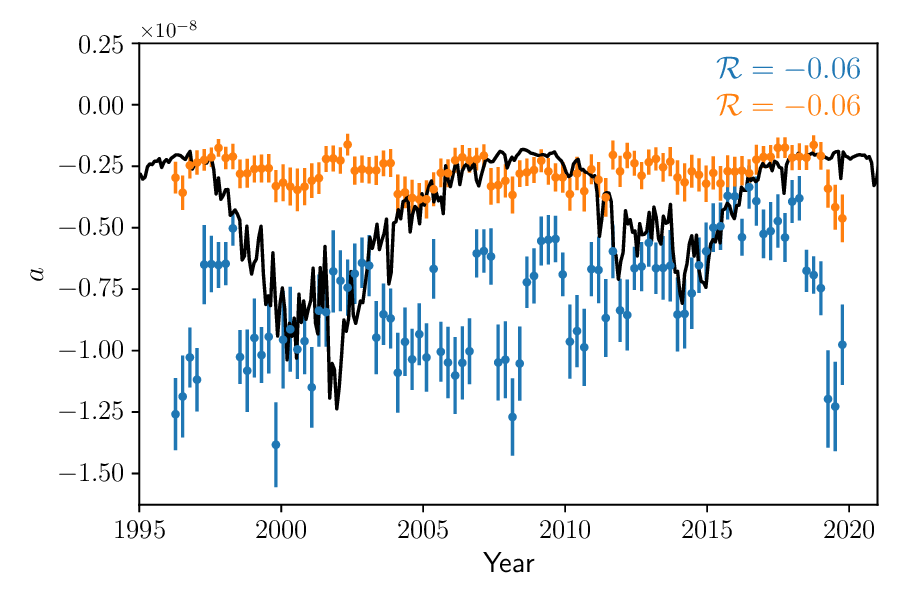} 
\end{subfigure}}
\resizebox{0.66\linewidth}{!}{
\begin{subfigure}[b]{.33\textwidth}
  \includegraphics[width=.99\linewidth]{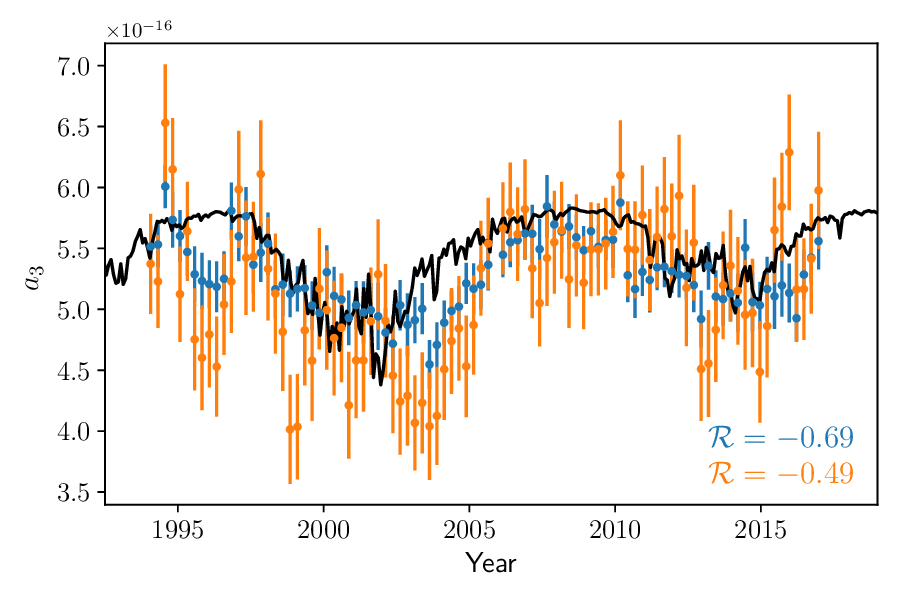}  
\end{subfigure}
\begin{subfigure}[b]{.33\textwidth}
  \includegraphics[width=.99\linewidth]{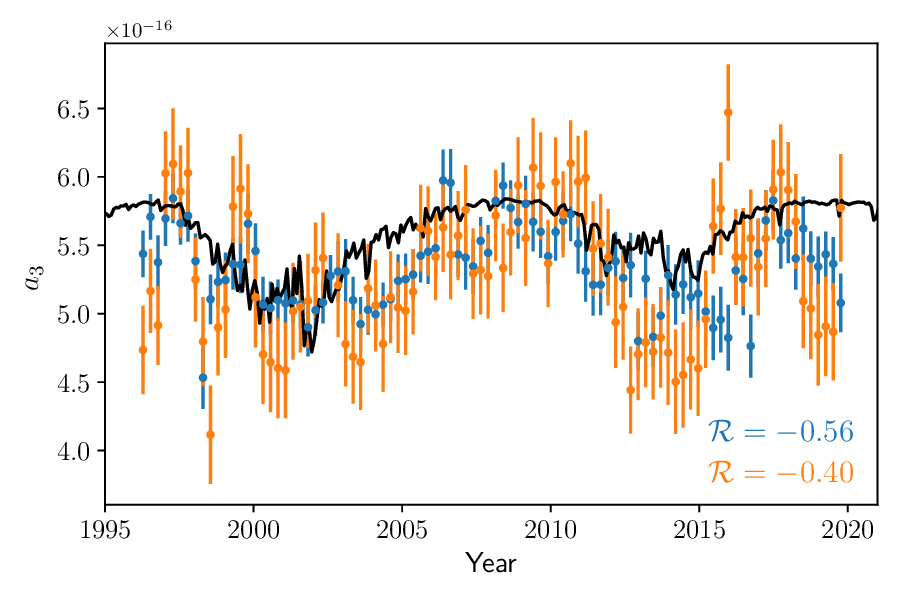}  
\end{subfigure}}
\caption{Imprint of magnetic activity cycle on the asteroseismic characterisation of the Sun. Two different datasets, receptively in blue ($n\geq 12$) and orange ($n\geq 16$), were investigated. The black line is the 10.7 cm radio emission flux, rescaled for illustration purposes and which serves as a proxy of the solar activity cycle. \textit{Upper line:} temporal evolution of the asteroseismic age. \textit{Lower panels:} temporal evolution of the free parameters of the two surface effect prescriptions that were investigated. \textit{From left to right:} impact of magnetic activity based on BiSON observations and using the BG2 surface effect prescription, GOLF observations and using the BG2 prescription, and GOLF observations and using the K1 prescription.}
\label{fig:pearsonr_age}
\end{figure*}

To maintain conciseness, we highlight the most notable findings here and provide a comprehensive table of all the correlations in Appendix~\ref{app:supplementary_data}. The results using the BG2 surface effect prescription are particularly relevant in the framework of the PLATO mission, as this prescription is widely adopted by the community and is considered the most robust
\citepalias[see e.g.][and references therein]{Betrisey2023_AMS_surf}. As illustrated in Fig.~\ref{fig:pearsonr_age}, we observe a discernible imprint of the activity cycle on the two free parameters of the BG2 prescription and the estimated stellar age across both GOLF and BiSON datasets. No imprint is found in the estimated solar mass, radius, and initial chemical composition. While the cycle's imprint on the surface prescription parameters indicates a partial absorption of activity effects, it is insufficient to prevent an impact on stellar age, contrary to literature expectations. Table~\ref{tab:variation_age} shows the age differences between solar minima and maxima of cycles 23 and 24. For cycle 23, we find variations of 5.8\% and 6.5\%, compared with the asteroseismic mean of the corresponding cycle, across both GOLF and BiSON datasets respectively, slightly reduced by 0.4\% and 0.9\% when including low-order frequencies. For the less active cycle 24, a smaller variation of 4.7\% is observed, as expected for lower activity levels. Nonetheless, these age variations are very significant compared to the 10\% age precision required by PLATO. Additionally, for the mode set with radial orders above $n=18$, the minimisation is less stable, leading to larger uncertainties that mask the activity cycle imprint. 

\begin{table}[t]
\caption{Age variation between solar minima and maxima of cycles 23 and 24.}
\centering
\begin{tabular}{lcccc}
\hline \hline
 & \multicolumn{2}{c}{cycle 23} & \multicolumn{2}{c}{cycle 24} \\ 
 & absolute & \% age & absolute & \% age \\
\hline 
\textit{GOLF} &  &  & & \\ 
BG2, $n\geq 12$ & 257 Myr & 5.4\% & 197 Myr & 4.1\% \\
BG2, $n\geq 16$ & 281 Myr & 5.8\% & 227 Myr & 4.7\% \\
K1, $n\geq 12$ & 264 Myr & 5.3\% & 299 Myr & 6.0\% \\
K1, $n\geq 16$ & 302 Myr & 6.1\% & 247 Myr & 5.0\% \\
\hline
\textit{BiSON} &  &  & & \\ 
BG2, $n\geq 12$ & 261 Myr & 5.4\% & 168 Myr & 3.5\% \\
BG2, $n\geq 16$ & 319 Myr & 6.5\% & 228 Myr & 4.7\% \\
\hline
\end{tabular} 
{\par\small\justify\textbf{Notes.} The left column represents the absolute variation and the right column compares the absolute variation with the mean asteroseismic age of the corresponding cycle. \par}
\label{tab:variation_age}
\end{table}

Comparing our results with those of \citet{Howe2017}, who demonstrated that magnetic activity's impact could be removed by filtering surface effects if the surface effects prescription parameters are the only free parameters, we note a key difference. We employed an advanced modelling procedure similar to what will be used in the PLATO pipeline, optimising additional parameters such as stellar mass, age, and initial chemical composition. Thus, it is not surprising that magnetic activity impacts other parameters than the free parameters of the surface effect prescription. Specifically, for the BG2 prescription, only the stellar age is affected, but not mass, radius, or initial chemical composition. On one hand, this could be due to the intrinsic nature of age in stellar models. The information contained in the oscillation frequencies provides a constraint on the stellar structure. Seismic data can therefore directly constrain stellar mass, radius, and initial composition. The stellar age however is a number that is associated with the stellar structure. It is  thus constrained by the seismic data in a more indirect way, and consequently, it is more sensitive to modelling inaccuracies. On the other hand, this could be due to the treatment of the stellar age during the interpolation process in AIMS. Further investigation with other minimisation softwares \citep[e.g. BASTA;][]{AguirreBorsen-Koch2022} would be relevant to clarify that aspect. Additionally, we also observe a constant age bias of about 300 Myr, primarily due to inaccuracies in the surface effect prescription and, to a lesser extent, the physical ingredients used in our models.

Using the BG2 prescription, we also detect a subtle imprint of the magnetic activity cycle on solar parameters such as the large separation, mean density, effective temperature, and absolute luminosity, with Pearson correlation coefficients ranging from 0.3 to 0.5 (see Table~\ref{tab:appendix_all_correlations} and Fig.~\ref{fig:appendix_pearsonr_age}). This observation was confirmed by smoothing the data with a Savitzky-Golay filter \citep{Savitzky&Golay1964}, and visually assessing that we can identify two distinct peaks corresponding to the cycle maxima. We note that these parameters are not free variables in our minimisation process. Thus, the observed imprint is primarily an indirect consequence of the effect on the optimised variables, particularly the stellar age. Notably, the weak imprint on the large separation aligns with existing literature, where such an impact has been documented for large separation values derived directly from observed solar frequencies \citep{Broomhall2011}.

When we apply the K1 surface effect prescription, the imprint of the magnetic activity cycle on the asteroseismic characterisation remains evident. However, in this case, the free parameter of the K1 prescription shows no correlation with the activity proxy. This suggests that the K1 correction cannot partially account for magnetic activity in the same way the BG2 prescription can. The K1 prescription is known to have limitations at high frequencies \citepalias[see e.g.][and references therein]{Betrisey2023_AMS_surf}, which are the most affected by magnetic activity, leading to difficulties in robustly estimating this parameter and decorrelating it from magnetic activity. Similar to the results with the BG2 prescription, we observe significant age variations of about 6\% between solar minima and maxima. Examining the non-optimised variables, we find several low yet non-negligible Pearson coefficients between 0.3 and 0.4 (see Table~\ref{tab:appendix_all_correlations}). However, the data smoothed with the Savitzky-Golay filter does not show clear excesses at the solar maxima, preventing us from conclusively identifying an imprint of magnetic activity in these variables (see Appendix~\ref{app:supplementary_data}). This does not imply that magnetic activity is negligible but rather that its impact is global, akin to a defect in physical ingredients, and suggests that the statistical uncertainty of the modelling procedure should be adjusted to account for this. Further investigation would be necessary to determine an appropriate quantitative correction. Nonetheless, given the lack of robustness of the K1 prescription, such efforts may not be justified.

Regarding the S1 prescription, we observed a bimodal distribution for the free parameter of the surface effect prescription in more than half of the minimisations. Knowing the expected values for the Sun, we could have discarded the unphysical solution, but doing so would introduce a bias inconsistent with the philosophy of our study. Consequently, we discarded the minimisations with bimodal distributions, which left us with too few data points to meaningfully compute the Pearson coefficient.


\section{Conclusions}
\label{sec:conclusions}
We carried out a detailed study of the impact of the magnetic activity cycle on the asteroseismic characterisation of the Sun-as-a-star based on 26 years of GOLF and BiSON Doppler velocity observations. In Sect.~\ref{sec:modelling_procedure}, we described the observational datasets of the oscillation frequencies and the `à la PLATO' modelling approach. The correlation of the solar parameters with the 10.7 cm radio emission flux, a proxy for the solar activity cycle, was then investigated in Sect.~\ref{sec:results}.

Our research has identified a clear impact of the solar magnetic activity cycle on the asteroseismic characterisation of the Sun, notably affecting the estimated solar age. This impact is evident across two independent datasets, GOLF and BiSON, and persists even when modifying the surface effect prescription. Consistently with literature predictions \citep{Howe2017}, we observed the cycle's influence on the two free parameters of the \citet{Ball&Gizon2014} surface effect prescription. While these parameters can therefore partially mitigate the effects of magnetic activity, they do not completely prevent its imprint on the solar parameters, contrary to previous expectations. Specifically, we found that the estimated solar age vary by about 6.2\% on average between solar minima and maxima, a substantial difference considering the 10\% age precision required by the PLATO mission for a Sun-like star. The variations are less pronounced, by about 0.8\% and 1.5\% on average, if low-order modes are included in the mode set or if the cycle is less active respectively. Using the \citet{Ball&Gizon2014} prescription, we also found a small imprint on the large separation, which is consistent with existing literature \citep{Broomhall2011}, the mean density, the effective temperature, and the absolute luminosity. Furthermore, the \citet{Kjeldsen2008} prescription proved ineffective in accounting for magnetic activity.

Considering future photometry missions such as PLATO, our study suggests that magnetic activity could present a substantial challenge. While Doppler velocity observations of the Sun-as-a-star by GOLF and BiSON deliver higher data quality than what is anticipated for the PLATO mission, it is important to note that the Sun is not particularly active \citep[e.g.][]{Reinhold2020,Santos2023}. In contrast, PLATO is expected to observe many more active stars, but with lower data quality, resulting in fewer detectable acoustic oscillations. Furthermore, these observations will likely span only a portion of the activity cycle \citep{Breton2024}, with time series insufficiently long to average out the effects of magnetic activity over one or several full cycles. Stellar characterisations may thus be influenced by the phase of the activity cycle, leading to potential biases if observations coincide with cycle extrema. Our study also raises questions about the necessity of re-evaluating the asteroseismic characterisation of the most active targets observed by \textit{Kepler}. This is an important endeavour since most methods for characterising planetary systems rely on the stellar characterisation. Thus, it is imperative to provide robust stellar parameters that would consistently account for magnetic activity. In future studies, we shall explore these issues further and investigate whether it is possible to mitigate the influence of magnetic activity using standard techniques employed to damp surface effects \citep[e.g.][]{Betrisey2023_AMS_surf}. Additionally, it would be worthwhile to examine the impact of magnetic activity on seismic inversion techniques, which also implicitly include magnetic activity within the parametrisation of the surface effects \citep[see e.g.][]{Pijpers2006,Betrisey&Buldgen2022,Buldgen2022c,Betrisey2023_AMS_surf,Betrisey2024_AMS_quality}.


\begin{acknowledgements}
J.B. acknowledges funding from the SNF AMBIZIONE and Postdoc.Mobility grants no. 185805 and no. P500PT{\_}222217. J.B. also thanks the Swiss Society for Astronomy and Astrophysics (SSAA) for covering travel expenses within the framework of this project. M.F. is a Postdoctoral Researcher of the Fonds de la Recherche Scientifique – FNRS. S.N.B acknowledges support from PLATO ASI-INAF agreement no. 2022-28-HH.0 "PLATO Fase D". R.A.G. acknowledges the support from PLATO and GOLF CNES grants. A.-M.B. has received support from STFC consolidated grant ST/T000252/1. A.M.A acknowledges support from the Swedish Research Council (VR 2020-03940) and from the Crafoord Foundation via the Royal Swedish Academy of Sciences (CR 2024-0015). O.K. acknowledges support by the Swedish Research Council (grant agreements no. 2019-03548 and 2023-03667), the Swedish National Space Agency, and the Royal Swedish Academy of Sciences.
\end{acknowledgements}


\bibliography{bibliography.bib}

\appendix
\section{Solar observational context}
\label{app:solar_observational_context}
Since the 1980s, global networks of ground-based telescopes, such as Interface Region Imaging Spectrograph \citep[IRIS;][]{Fossat1988}, Birmingham Solar Oscillations Network \citep[BiSON;][]{Davies2014_bison,Hale2016}, Global Oscillations Network Group \citep[GONG;][]{Harvey1996}, and Stellar Observations Network Group \citep[SONG;][]{Grundahl2006} have been monitoring these oscillations at high cadence and high temporal resolution. For the SONG network, it should be noted that the solar component of the network is called Solar-SONG \citep[see e.g.][and references therein]{Breton2022_helio}. Additionally, space-based observations have been made since the mid-1990s by the Variability of solar IRradiance and Gravity Oscillations / Sun PhotoMeters \citep[VIRGO/SPM;][]{Frohlich1995}, Michelson Doppler Imager \citep[MDI;][]{Scherrer1995}, and Global Oscillations at Low Frequencies \citep[GOLF;][]{Gabriel1995} instruments on board of the Solar and Heliospheric Observatory \citep[SoHO;][]{Domingo1995}. Since the early 2010s, the Sun is also monitored by the Helioseismic and Magnetic Imager \citep[HMI;][]{Scherrer2012} on board of the Solar Dynamics Observatory \citep[SDO;][]{Pesnell2012}. This unique dataset provides an excellent opportunity to study the effects of stellar activity, with high-quality acoustic oscillation data being collected continuously for over 30 years, and fully covering solar cycles 23 and 24.


\section{Detailed modelling procedure}
\label{app:detailed_modeling_procedure}
The GOLF observations were divided into 94 yearly overlapping snapshots, each delayed by 91.25 days, starting from April 11, 1996, and concluding on July 6, 2020. The acoustic frequency of the p modes in GOLF yearly time series were extracted with the Markov Chain Monte Carlo (MCMC) asteroseismic module \textsc{apollinaire} \citep{Breton2022_astro,Breton2022_helio} by analysing the time series power spectral density (PSD) with the following strategy. After removing the background, p modes were fitted by pair $\ell = 0, 2$ and $1, 3$, considering an asymmetric Lorentzian profile. Mode height, width, and rotational splitting was fitted independently for each mode, while asymmetry parameter was taken to be common for both modes of a given pair. Power leakage from intermediate-degree $\ell=4,5$ was accounted for. The MCMC sampling procedure used the \textsc{emcee} \citep{Foreman-Mackey2013} ensemble sampler. Chains were sampled using 500 walkers and 1000 steps, with the 400 first steps discarded as burnt-in. 

Similarly, the BiSON observations were divided into 92 overlapping snapshots, beginning on December 30, 1993, and ending on September 23, 2016. It should be noted that the BiSON time series are publicly available at the BiSON Open Data Portal\footnote{ \url{http://bison.ph.bham.ac.uk/portal/timeseries}}. The acoustic oscillation frequencies were then extracted using the fitting procedure described in \citet{Fletcher2009}.

For each snapshot, the fundamental solar parameters (e.g. mass, radius, age, etc. See Table \ref{tab:appendix_all_correlations} for the complete list) were determined using an advanced `à la PLATO' modelling procedure. To this end, the acoustic frequencies and non-seismic constraints (in our case, the spectroscopic constraints: effective temperature, metallicity, and luminosity) were fitted using the AIMS software \citep{Rendle2019}. As non-seismic constraints, we adopted $T_{\mathrm{eff}}=5772\pm 85$ K \citep{IAU2015-ResolutionB3}, $\rm [Fe/H] = 0.00\pm 0.10$, and $L=1.00\pm 0.03~L_\odot$. The uncertainties of the non-seismic constraints were adjusted to match the data quality of the best \textit{Kepler} targets. AIMS, which is an MCMC-based algorithm, is based on the \textsc{emcee} package and employs a Bayesian approach to provide posterior probability distributions of the optimised stellar parameters. AIMS also incorporates an interpolation scheme to sample between grid points. We used the standard MS subgrid of the \textit{Spelaion} grid from \citet{Betrisey2023_AMS_surf}. The combination of this high-resolution grid and the interpolation scheme allows for thorough exploration of the parameter space. Four main free parameters (mass, age, and initial hydrogen and helium mass fractions $X_0$ and $Y_0$) were optimised, along with one or two additional free parameters depending on the surface effect prescription considered (one for the K1 and S1 surface effect prescriptions and two for the BG2 prescription). Uniform `uninformative' priors were applied to the estimated stellar parameters, except for the stellar age, for which we employed a uniform distribution in the interval [0, 13.8] Gyr, and likelihoods were calculated under the assumption that the true observational values were perturbed by normally distributed random noise.


\section{Detailed evaluation procedure of the Pearson coefficient}
\label{app:supplementary_data}
To evaluate the correlation between solar parameters and the magnetic activity cycle, we computed the Pearson correlation coefficient $\mathcal{R}$ \citep{Pearson1895} using the 10.7 cm radio emission flux, a well-established proxy for solar activity \citep[see e.g.][and references therein]{Tapping2013}. The 10.7 cm flux data, which is recorded daily in Canada since 1947 and nowadays even includes three measurements per day\footnote{see \url{https://www.spaceweather.gc.ca/}}, was smoothed to a monthly average. Since the GOLF and BiSON data points do not exactly match the dates of the monthly flux, we linearly interpolated the flux values to ensure consistent calculation of the Pearson coefficient. Additionally, due to the construction of the GOLF and BiSON acoustic oscillation datasets, where data points are correlated over four consecutive points, we created uncorrelated datasets by selecting one data point out of every four. This process resulted in four subsets of the data. For each subset, we calculated the Pearson correlation coefficient. The final Pearson coefficient reported in the tables and figures of this article is the mean of the coefficients from these four subsets. The standard deviation of these coefficients provides an uncertainty measure, reflecting data sensitivity and the confidence level in the estimated Pearson coefficient. It is worth noting that ignoring the correlations in the construction of GOLF and BiSON acoustic oscillation datasets yields similar Pearson coefficients. This similarity is expected, as the construction of these datasets only introduces minor correlations. We provide in Table~\ref{tab:appendix_all_correlations} the Pearson correlation coefficient between the solar parameters and the 10.7 cm radio emission flux for the different configurations that were investigated in our study. 

\begin{figure*}
\resizebox{0.999\linewidth}{!}{
\begin{subfigure}[b]{.49\textwidth}
  \includegraphics[width=.99\linewidth]{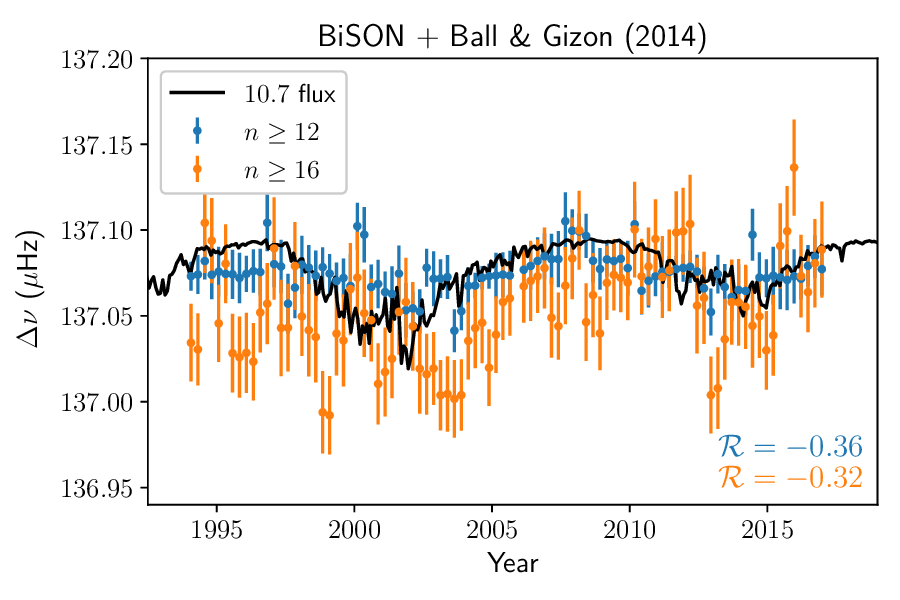}  
\end{subfigure}
\begin{subfigure}[b]{.49\textwidth}
  \includegraphics[width=.99\linewidth]{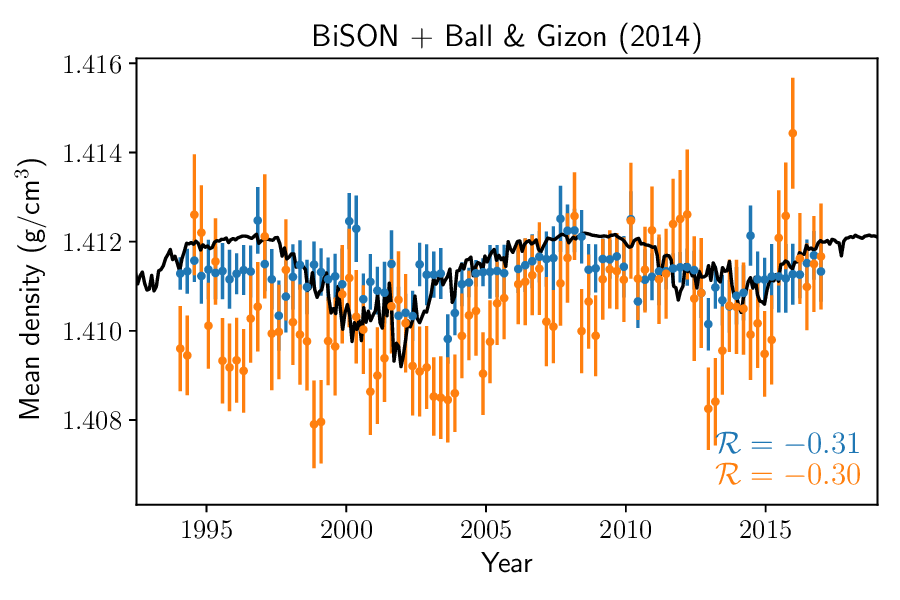} 
\end{subfigure}}
\resizebox{0.999\linewidth}{!}{
\begin{subfigure}[b]{.49\textwidth}
  \includegraphics[width=.99\linewidth]{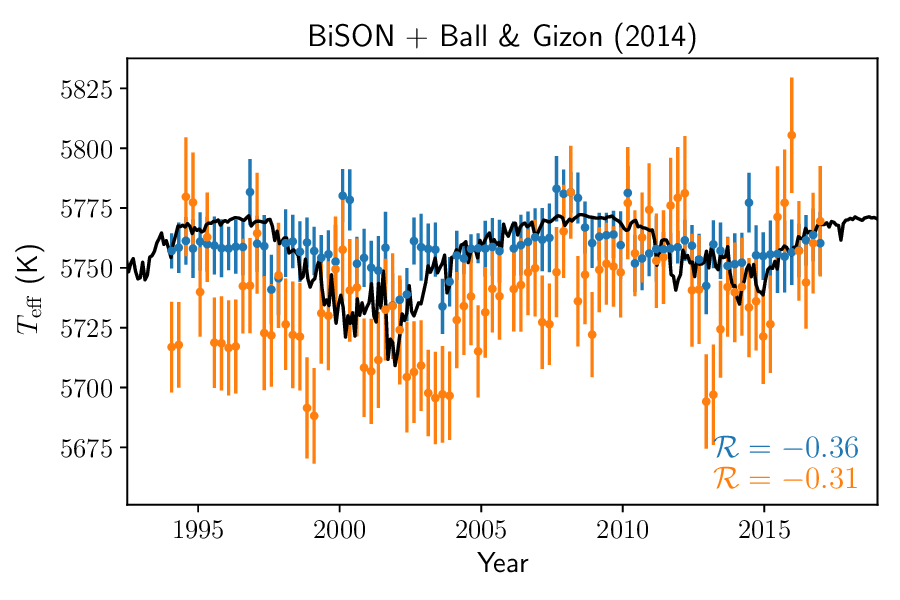}  
\end{subfigure}
\begin{subfigure}[b]{.49\textwidth}
  \includegraphics[width=.99\linewidth]{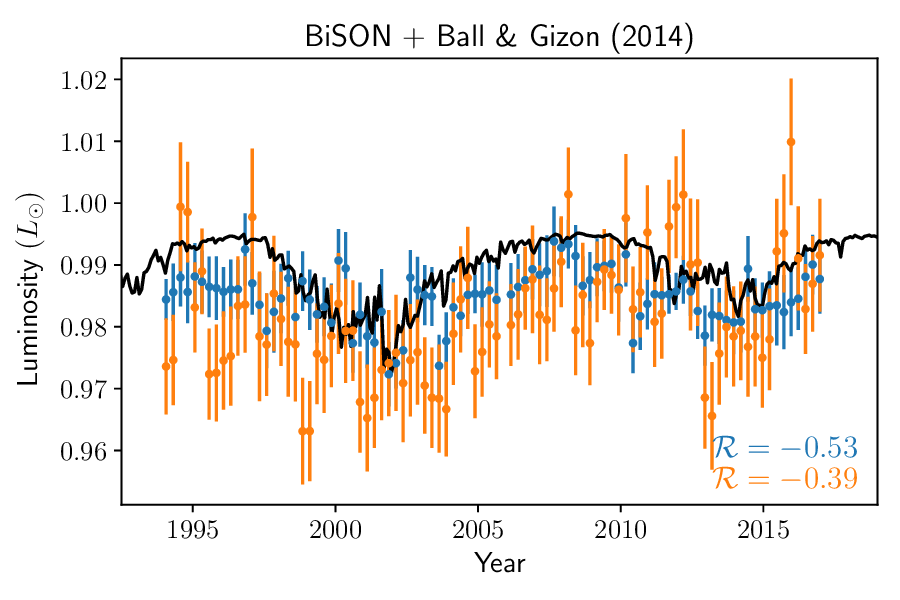} 
\end{subfigure}}
\resizebox{0.999\linewidth}{!}{
\begin{subfigure}[b]{.49\textwidth}
  \includegraphics[width=.99\linewidth]{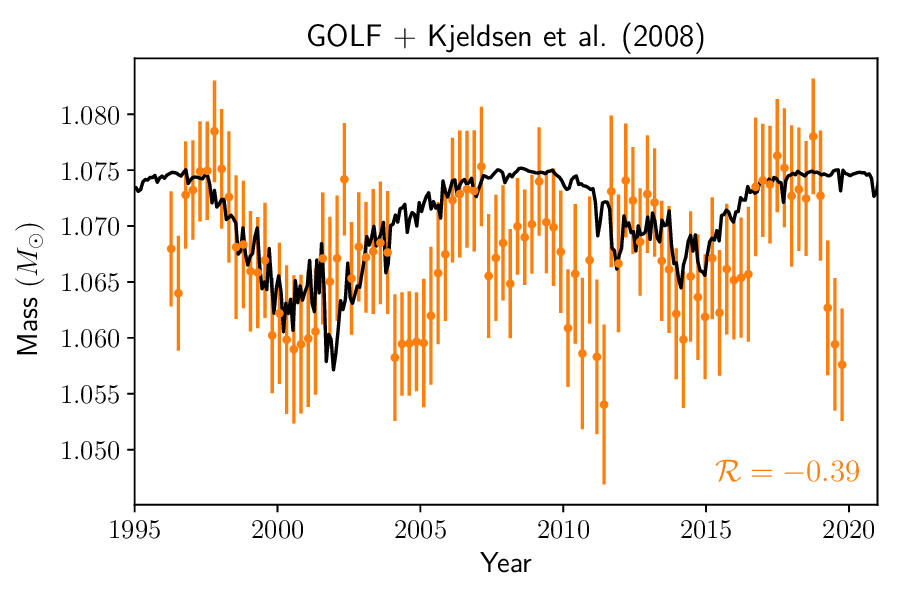}  
\end{subfigure}
\begin{subfigure}[b]{.49\textwidth}
  \includegraphics[width=.99\linewidth]{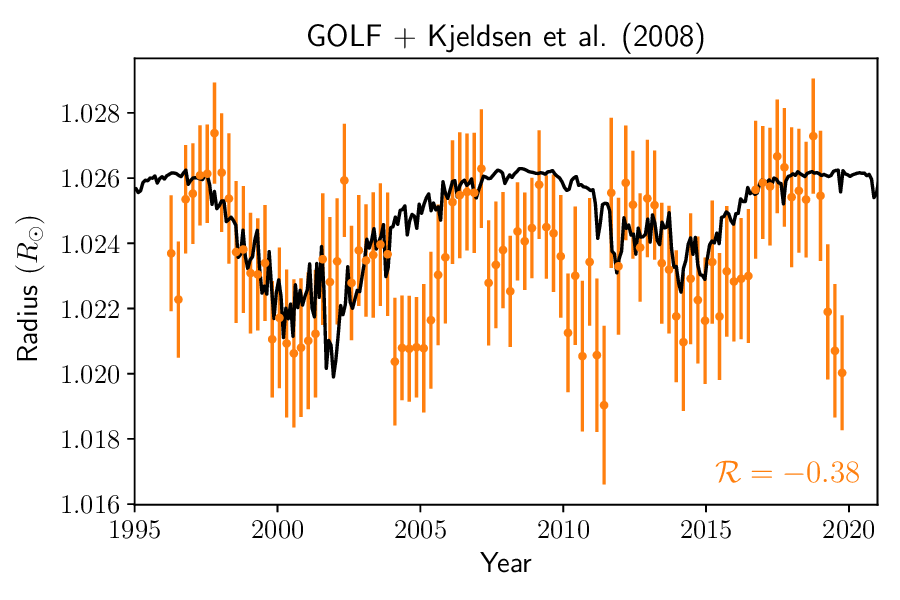}  
\end{subfigure}}
\caption{Imprint of magnetic activity cycle on the large separation, mean density, effective temperature, and absolute luminosity using the \citet{Ball&Gizon2014} surface effect prescription, and on the solar mass and radius using the \citet{Kjeldsen2008} prescription.}
\label{fig:appendix_pearsonr_age}
\end{figure*}

In Fig.~\ref{fig:appendix_pearsonr_age}, we show the imprint of magnetic activity cycle on the large separation, mean density, effective temperature, and absolute luminosity using the \citet{Ball&Gizon2014} surface effect prescription, and on the solar mass and radius using the \citet{Kjeldsen2008} prescription. For the imprints with \citet{Ball&Gizon2014} prescription, we smoothed the data with a Savitzky-Golay filter, and confirmed visually that we can clearly identify two distinct peaks corresponding to the cycle maxima. For the \citet{Kjeldsen2008} prescription, we did not find a clear imprint of the cycle by visually inspecting the smoothed data. 

\begin{table*}
\centering
\caption{Pearson correlation coefficient between the solar parameters and the 10.7 cm radio emission flux for the different configurations that were investigated in this study.}
\resizebox{\linewidth}{!}{
\begin{tabular}{lccccccccc}
\hline \hline
 & \multicolumn{3}{c}{BiSON + BG2} & \multicolumn{3}{c}{GOLF + BG2} & \multicolumn{3}{c}{GOLF + K1} \\
 & cycle 23 & cycle 24 & two cycles & cycle 23 & cycle 24 & two cycles & cycle 23 & cycle 24 & two cycles \\
\hline
$n \geq 12$ & & & & & & & & & \\
Mass & $-0.04 \pm 0.21$ & $0.10 \pm 0.30$ & $-0.01 \pm 0.22$ & $0.05 \pm 0.28$ & $0.18 \pm 0.32$ & $0.09 \pm 0.21$ & $-0.23 \pm 0.10$ & $-0.11 \pm 0.21$ & $-0.31 \pm 0.08$ \\ 
$Y_0$ & $-0.09 \pm 0.22$ & $-0.12 \pm 0.41$ & $-0.11 \pm 0.21$ & $-0.13 \pm 0.31$ & $-0.18 \pm 0.30$ & $-0.13 \pm 0.21$ & $0.19 \pm 0.17$ & $0.11 \pm 0.19$ & $0.29 \pm 0.06$ \\ 
$\log Z_0$ & $0.14 \pm 0.18$ & $0.04 \pm 0.24$ & $0.11 \pm 0.15$ & $0.17 \pm 0.12$ & $0.11 \pm 0.10$ & $0.06 \pm 0.06$ & $0.03 \pm 0.07$ & $0.07 \pm 0.20$ & $-0.05 \pm 0.08$ \\ 
Age & $0.64 \pm 0.13$ & $0.64 \pm 0.11$ & $0.65 \pm 0.12$ & $0.57 \pm 0.12$ & $0.43 \pm 0.12$ & $0.50 \pm 0.10$ & $0.69 \pm 0.10$ & $0.32 \pm 0.14$ & $0.40 \pm 0.02$ \\ 
$a_{-1}$ & $0.77 \pm 0.09$ & $0.72 \pm 0.09$ & $0.77 \pm 0.07$ & $0.70 \pm 0.05$ & $0.63 \pm 0.21$ & $0.63 \pm 0.10$ & - & - & - \\ 
$a_3$ & $-0.69 \pm 0.11$ & $-0.65 \pm 0.08$ & $-0.69 \pm 0.07$ & $-0.58 \pm 0.07$ & $-0.63 \pm 0.20$ & $-0.56 \pm 0.09$ & - & - & - \\ 
$a$ & - & - & - & - & - & - & $0.13 \pm 0.12$ & $0.09 \pm 0.27$ & $-0.06 \pm 0.11$ \\ 
Radius & $-0.02 \pm 0.21$ & $0.14 \pm 0.29$ & $0.02 \pm 0.22$ & $0.07 \pm 0.27$ & $0.19 \pm 0.32$ & $0.09 \pm 0.21$ & $-0.21 \pm 0.10$ & $-0.11 \pm 0.22$ & $-0.30 \pm 0.08$ \\ 
Mean density & $-0.31 \pm 0.22$ & $-0.37 \pm 0.28$ & $-0.31 \pm 0.21$ & $-0.27 \pm 0.10$ & $-0.27 \pm 0.29$ & $-0.17 \pm 0.17$ & $0.03 \pm 0.09$ & $0.01 \pm 0.24$ & $0.17 \pm 0.08$ \\ 
$\log g$ & $-0.10 \pm 0.20$ & $0.02 \pm 0.34$ & $-0.06 \pm 0.22$ & $0.01 \pm 0.30$ & $0.16 \pm 0.31$ & $0.07 \pm 0.21$ & $-0.25 \pm 0.11$ & $-0.13 \pm 0.21$ & $-0.32 \pm 0.08$ \\ 
$T_{\mathrm{eff}}$ & $-0.33 \pm 0.20$ & $-0.45 \pm 0.27$ & $-0.36 \pm 0.19$ & $-0.30 \pm 0.12$ & $-0.28 \pm 0.30$ & $-0.22 \pm 0.18$ & $-0.03 \pm 0.10$ & $0.00 \pm 0.23$ & $0.15 \pm 0.08$ \\ 
$\rm [Fe/H]$ & $0.10 \pm 0.16$ & $-0.03 \pm 0.20$ & $0.06 \pm 0.11$ & $0.12 \pm 0.19$ & $0.06 \pm 0.06$ & $0.02 \pm 0.05$ & $0.06 \pm 0.11$ & $0.09 \pm 0.19$ & $0.01 \pm 0.08$ \\ 
Luminosity & $-0.52 \pm 0.18$ & $-0.51 \pm 0.27$ & $-0.53 \pm 0.16$ & $-0.43 \pm 0.04$ & $-0.31 \pm 0.23$ & $-0.30 \pm 0.14$ & $-0.22 \pm 0.11$ & $-0.07 \pm 0.24$ & $0.02 \pm 0.07$ \\ 
$\nu_{\mathrm{max}}$ & $0.07 \pm 0.21$ & $0.22 \pm 0.27$ & $0.10 \pm 0.22$ & $0.14 \pm 0.23$ & $0.21 \pm 0.32$ & $0.13 \pm 0.20$ & $-0.15 \pm 0.10$ & $-0.08 \pm 0.22$ & $-0.26 \pm 0.08$ \\ 
$\Delta\nu$ & $-0.36 \pm 0.22$ & $-0.40 \pm 0.28$ & $-0.36 \pm 0.21$ & $-0.28 \pm 0.13$ & $-0.28 \pm 0.30$ & $-0.20 \pm 0.17$ & $0.01 \pm 0.09$ & $0.01 \pm 0.23$ & $0.17 \pm 0.08$ \\ 
\hline
$n \geq 16$ & & & & & & & & & \\
Mass & $0.02 \pm 0.12$ & $0.14 \pm 0.07$ & $0.16 \pm 0.02$ & $0.02 \pm 0.16$ & $0.37 \pm 0.06$ & $0.16 \pm 0.07$ & $-0.45 \pm 0.13$ & $-0.34 \pm 0.30$ & $-0.39 \pm 0.19$ \\ 
$Y_0$ & $-0.09 \pm 0.11$ & $-0.26 \pm 0.17$ & $-0.23 \pm 0.05$ & $-0.14 \pm 0.22$ & $-0.36 \pm 0.06$ & $-0.22 \pm 0.14$ & $0.26 \pm 0.14$ & $0.30 \pm 0.32$ & $0.20 \pm 0.23$ \\ 
$\log Z_0$ & $0.10 \pm 0.04$ & $-0.11 \pm 0.40$ & $0.04 \pm 0.16$ & $-0.06 \pm 0.32$ & $0.19 \pm 0.20$ & $0.05 \pm 0.21$ & $-0.29 \pm 0.14$ & $-0.23 \pm 0.34$ & $-0.29 \pm 0.17$ \\ 
Age & $0.66 \pm 0.06$ & $0.58 \pm 0.07$ & $0.67 \pm 0.05$ & $0.66 \pm 0.08$ & $0.61 \pm 0.09$ & $0.60 \pm 0.05$ & $0.76 \pm 0.05$ & $0.35 \pm 0.05$ & $0.57 \pm 0.09$ \\ 
$a_{-1}$ & $0.58 \pm 0.11$ & $0.43 \pm 0.06$ & $0.55 \pm 0.02$ & $0.43 \pm 0.03$ & $0.55 \pm 0.07$ & $0.48 \pm 0.02$ & - & - & - \\ 
$a_3$ & $-0.51 \pm 0.12$ & $-0.34 \pm 0.09$ & $-0.49 \pm 0.04$ & $-0.31 \pm 0.05$ & $-0.52 \pm 0.10$ & $-0.40 \pm 0.03$ & -& - & - \\ 
$a$ & - & - & - & - & - & - & $-0.02 \pm 0.11$ & $-0.12 \pm 0.36$ & $-0.06 \pm 0.17$ \\ 
Radius & $0.04 \pm 0.12$ & $0.15 \pm 0.07$ & $0.17 \pm 0.02$ & $0.03 \pm 0.15$ & $0.37 \pm 0.07$ & $0.17 \pm 0.06$ & $-0.44 \pm 0.13$ & $-0.34 \pm 0.30$ & $-0.38 \pm 0.19$ \\ 
Mean density & $-0.33 \pm 0.13$ & $-0.18 \pm 0.20$ & $-0.30 \pm 0.05$ & $-0.12 \pm 0.11$ & $-0.39 \pm 0.13$ & $-0.23 \pm 0.07$ & $0.28 \pm 0.14$ & $0.26 \pm 0.33$ & $0.27 \pm 0.18$ \\ 
$\log g$ & $-0.02 \pm 0.12$ & $0.12 \pm 0.07$ & $0.13 \pm 0.03$ & $-0.00 \pm 0.17$ & $0.35 \pm 0.05$ & $0.15 \pm 0.08$ & $-0.48 \pm 0.14$ & $-0.35 \pm 0.29$ & $-0.41 \pm 0.19$ \\ 
$T_{\mathrm{eff}}$ & $-0.30 \pm 0.10$ & $-0.17 \pm 0.17$ & $-0.31 \pm 0.05$ & $-0.18 \pm 0.12$ & $-0.45 \pm 0.12$ & $-0.28 \pm 0.08$ & $0.14 \pm 0.12$ & $0.21 \pm 0.33$ & $0.18 \pm 0.17$ \\ 
$\rm [Fe/H]$ & $0.05 \pm 0.06$ & $-0.14 \pm 0.41$ & $-0.01 \pm 0.17$ & $-0.09 \pm 0.33$ & $0.12 \pm 0.21$ & $0.00 \pm 0.22$ & $-0.30 \pm 0.14$ & $-0.23 \pm 0.34$ & $-0.29 \pm 0.17$ \\ 
Luminosity & $-0.50 \pm 0.08$ & $-0.14 \pm 0.27$ & $-0.39 \pm 0.08$ & $-0.28 \pm 0.17$ & $-0.46 \pm 0.14$ & $-0.33 \pm 0.13$ & $-0.00 \pm 0.11$ & $0.14 \pm 0.33$ & $0.07 \pm 0.14$ \\ 
$\nu_{\mathrm{max}}$ & $0.09 \pm 0.11$ & $0.16 \pm 0.09$ & $0.20 \pm 0.03$ & $0.07 \pm 0.14$ & $0.41 \pm 0.08$ & $0.20 \pm 0.06$ & $-0.32 \pm 0.13$ & $-0.29 \pm 0.32$ & $-0.30 \pm 0.19$ \\ 
$\Delta\nu$ & $-0.34 \pm 0.13$ & $-0.20 \pm 0.17$ & $-0.32 \pm 0.04$ & $-0.15 \pm 0.08$ & $-0.40 \pm 0.11$ & $-0.25 \pm 0.06$ & $0.21 \pm 0.15$ & $0.28 \pm 0.31$ & $0.24 \pm 0.17$ \\ 
\hline
\end{tabular}
}
\label{tab:appendix_all_correlations}
\end{table*}

\end{document}